\documentstyle[12pt]{article}
\begin{document}
\begin{center}
{\Large Spin polarised magnetized cylinders in torsioned spacetime}
\vspace{1cm}
\noindent

L.C.Garcia de Andrade\footnote{Departamento de Fisica Teorica,Instituto de F\'{\i}sica , UERJ, Rua S\~{a}o
francisco Xavier 524, Rio de Janeiro,CEP:20550-013, Brasil.e-mail:garcia@dft.if.uerj.br.}
\end{center} 
\vspace{2cm}
\begin{center}
{\Large Abstract}
\end{center}
\vspace{0.5cm}
A Spin-polarised cylindrically symmetric exact class of solutions endowed with magnetic fields in Einstein-Cartan-Maxwell gravity is obtained. Application of matching conditions to this interior solution having an exterior as Einstein's vacuum solution shows that for this class of metrics the Riemann-Cartan (RC) rotation vanishes which makes the solution static. Therefore we end up with a magnetized static spin polarised cylinder where the pressure along the symmetry axis is negative.   
\newpage
Recently we showed that spin-polarised cylinders in the EC theory of gravity \cite{1,2,3} maybe used with the purposes of torsion detection \cite{4}. In this letter we show that Soleng \cite{3} cylinder geometry given by
\begin{equation}
ds^{2}=-(e^{\alpha}dt+Md{\phi})^{2}+r^{2}e^{-2{\alpha}}d{\phi}^{2}+e^{2{\beta}-2{\alpha}}(dr^{2}+dz^{2}).
\label{1}
\end{equation}
in the particular case of ${\alpha}={\beta}=0$ can be shown to be a solution of Einstein-Cartan-Maxwell (ECM) field equations which represents a magnetized spin polarised cylinder in torsioned spacetime where the RC rotation vanishes when one applies the matching condition in this non-Riemannian space. Here $M$ is a function of the radial coordinate $r$. The exterior solution is the same as used by Soleng in the case of thick spinning cosmic strings \cite{5} and represents an exterior solution of Einstein's vacuum field equation
\begin{equation}
R_{ik}=0
\label{2}
\end{equation}
$(i,j=0,1,2,3)$ as
\begin{equation}
ds^{2}=-dt^{2}-2adtd{\phi}+dr^{2}+(B^{2}(r+r_{0})^{2}-a^{2})d{\phi}^{2}+dz^{2}
\label{3}
\end{equation}
where $a$,B and $r_{0}$ are constants. Before proceed in this analysis let us consider the above metric (\ref{1}) in terms of the differential one-form basis 
\begin{equation} 
{\theta}^{0}=e^{\alpha}dt+Md{\phi} , 
\label{4} 
\end{equation}
\begin{equation}
{\theta}^{1}=e^{{\beta}-{\alpha}}dr ,
\label{5}
\end{equation}
\begin{equation}
{\theta}^{2}=re^{-{\alpha}}d{\phi} ,
\label{6}
\end{equation}
\begin{equation}
{\theta}^{3}=e^{{\beta}-{\alpha}}dz .
\label{7}
\end{equation}
Polarisation along the axis of symmetry is considered and the Cartan torsion is given in terms of differential forms by
\begin{equation}
T^{i}=2k{\sigma}{\delta}^{i}_{0}{\theta}^{1}{\wedge}{\theta}^{2}
\label{8}
\end{equation}
where ${\sigma}$ is a constant spin density. For computational convenience we addopt Soleng's definition \cite{3} for the RC rotation ${\Omega}$
\begin{equation}
{\Omega}:= -\frac{1}{2}{\sigma}+\frac{M'}{2r}
\label{9}
\end{equation}
where ${\Omega}$ is the cylinder RC vorticity. Cartan's first structure equation is  
\begin{equation}
T^{i}=d{\theta}^{i}+{{\omega}^{i}}_{k}{\wedge}{\theta}^{k}
\label{10}
\end{equation}
and determines the connection forms ${{\omega}^{i}}_{j}$.The conection one-forms are given by
\begin{equation}
{\omega}^{0}_{1}= -{\Omega}{\omega}^{2}
\label{11}
\end{equation}
\begin{equation}
{\omega}^{0}_{2}={\Omega}{\omega}^{1}
\label{12}
\end{equation}
\begin{equation}
{\omega}^{0}_{3}=0
\label{13}
\end{equation}
\begin{equation}
{\omega}^{1}_{2}=-{\Omega}{\omega}^{0}-(\frac{1}{r}){\omega}^{2}
\label{14}
\end{equation}
while others vanish. From the Cartan's second structure equation
\begin{equation}
{R^{i}}_{j}=d{{\omega}^{i}}_{j}+{{\omega}^{i}}_{k}{\wedge}{{\omega}^{k}}_{j}
\label{15}
\end{equation}
where the curvature RC forms ${R^{i}}_{j}={R^{i}}_{jkl}{\theta}^{k}{\wedge}{\theta}^{l}$ where ${R^{i}}_{jkl}$ is the RC curvature tensor. This is accomplished by computing the RC curvature components from the Cartan structure equations as
\begin{equation}
R_{0101}={\Omega}^{2} ,
\label{16}
\end{equation}
\begin{equation}
R_{0112}={\Omega}' ,
\label{17}
\end{equation}
\begin{equation}
R_{0202}={\Omega}^{2} ,
\label{18}
\end{equation}
\begin{equation}
R_{1201}={\Omega}' ,
\label{19}
\end{equation}
\begin{equation}
R_{1212}=3{\Omega}^{2}-2{\Omega}{\sigma} ,
\label{20}
\end{equation}
others zero. The dash here represents the derivative $w.r.t$ to the radial coordinate $r$. From the curvature expressions above it is possible to built the ECM field equations as
\begin{equation}
{G^{i}}_{k}= k{T^{i}}_{k}
\label{21}
\end{equation}
where ${G^{i}}_{k}$ is the Einstein-Cartan tensor and ${T^{i}}_{k}$ is the total energy-momentum tensor composed of the fluid tensor ${T^{i}}_{k}=({\rho},p_{r},0,p_{z})$ and the electromagnetic field tensor 
\begin{equation}
{t^{i}}_{k}= ({F^{i}}_{l} {F^{l}}_{k}-\frac{1}{2}{\delta}^{i}_{k}(E^{2}-B^{2}))
\label{22}
\end{equation}
where $F_{0{\gamma}}$ correspond to the electric field $\vec{E}$ while $F_{{\alpha}{\beta}}$ components of the Maxwell tensor field $F_{ij}$ correspond to the magnetic field $\vec{B}$. Here we consider that the electric field vanishes along the cylinder, and ${{\alpha}=1,2,3}$. Thus the natural notation $E^{2}= (\vec{E})^{2}$ and the same is valid for the magnetic field. Thus explicitly the ECM equations read
\begin{equation}
-3{\Omega}^{2}-{\sigma}{\Omega}=-k({\rho}+\frac{B_{z}}{2})
\label{23}
\end{equation}
\begin{equation}
{\Omega}^{2}=k(p_{r}-\frac{B_{z}}{2})
\label{24}
\end{equation}
\begin{equation}
{\Omega}'=0
\label{25}
\end{equation}
\begin{equation}
{\Omega}^{2}+{\sigma}{\Omega}=-k(p_{z}+\frac{B_{z}}{2})
\label{26}
\end{equation}
Note that equation (\ref{25}) is the simplest to solve and yields ${\Omega}={\Omega}_{0}= constant$. Therefore so far just from the ECM field equations we cannot say that the cylinder is static. Before proceed therefore it is useful to show that this results from the Arkuszewski-Kopczynski-Ponomariev (AKP) \cite{6} junction conditions for Einstein-Cartan gravity which match an interior solution of ECM field equations to the exterior vacuum solution given by the geometry given by expression (\ref{3}). The AKP conditions are
\begin{equation}
{g}_{ij,r}|_{+}=g_{ij,r}|_{-} - 2K_{r(ij)}
\label{27}
\end{equation}
for $(i,j)$ distinct from the r coordinate, where the contortion tensor is
\begin{equation}
K_{ijk}= \frac{1}{2}( T_{jik}+T_{jki}-T_{ijk} )
\label{28}
\end{equation}
where $T_{jik}$ is the Cartan torsion. The plus and minus signs here correspond respectively to the exterior and interior spacetimes respectively. The others AKP conditions state that the fluid elements do not move across the junction surface, the stress normal to the junction surface vanishes and that 
\begin{equation}
{g}_{ij}|_{+}=g_{ij}|_{-} 
\label{29}
\end{equation}
which is the general relativistic Lichnerowicz condition. From the cylinder geometry one obtains 
\begin{equation}
g_{t{\phi}}|_{+}= g_{t{\phi}}|_{-}
\label{30}
\end{equation}
\begin{equation}
{g}_{{\phi}{\phi}}|_{+}=g_{{\phi}{\phi}}|_{-} 
\label{31}
\end{equation}
\begin{equation}
g_{t{\phi},r}|_{+}= g_{t{\phi},r}|_{-} - T_{tr{\phi}}
\label{32}
\end{equation}  
\begin{equation}
g_{{\phi}{\phi},r}|_{+}= g_{{\phi}{\phi},r}|_{-} - 2 T_{{\phi}r{\phi}}
\label{33}
\end{equation}
which for the exrerior and interior of the cylinder matching at $r=R$ one obtains 
\begin{equation}
a=M(R) 
\label{34}
\end{equation}
\begin{equation}
B^{2}(R+r_{0})^{2}= R^{2}- M^{2} 
\label{35}
\end{equation} 
\begin{equation}
{\Omega}_{0}= -\frac{1}{2}{\sigma}_{0}+\frac{M'}{2R} 
\label{36}
\end{equation} 
\begin{equation}
0= {\sigma}_{0}R-M' 
\label{37}
\end{equation}  
\begin{equation}
B^{2}(R+r_{0})= R -MM'+MR{\sigma}_{0} 
\label{38}
\end{equation}
Substitution of $M'$ above into expression (\ref{36}) yields the desired result that the Riemann-Cartan rotation ${\Omega}$ vanishes. The remaining junction conditions \cite{7} yield
\begin{equation}
B^{2}=\frac{R}{(R+r_{0})} 
\label{39}
\end{equation}
\begin{equation}
{\sigma}_{0}= \frac{4}{R}[1-\frac{R+r_{0}}{R}]
\label{40}
\end{equation}
Substitution of these results into the exterior metric yields the following exterior spacetime for the spin polarised cylinder 
\begin{equation}
ds^{2}=-dt^{2}-2{\sigma}_{0}R^{2}dtd{\phi}+dr^{2}+R(\frac{(r+r_{0})^{2}}{R+r_{0}}-{{\sigma}_{0}}^{2}R^{3})d{\phi}^{2}+dz^{2}
\label{41}
\end{equation}
Now going back to the ECM equations we obtain the following constraints
\begin{equation}
{\rho}=\frac{B_{z}}{2} 
\label{42}
\end{equation}
which states that the energy density is purely of magnetic origin. Besides to keep the stability of the spin polarised cylinder and its static nature one obtains that the radial pressure $p_{r}>0$ while the axial pressure is negative or $p_{z}<0$. Indeed from the field equations we obtain $p_{z}= - \frac{{B_{z}}^{2}}{2}$ while $p_{r}= \frac{{B_{z}}^{2}}{2}$. Physically this is in accordance with the fact that the radial stresses , here including the magnetic stress $(T^{1}_{1})$ must vanish at the cylinder surface. The heat flow also vanishes which in the ECM field equations implies that the RC rotation must be in principle constant. Physical applications of the model discuss here may be in the investigation of the gravitational extra effects on the well-known Einstein-de Haas effect due to the non-Riemannian effects from the spin density. Unfortunatly due to the fact that the magnetic field is strong for the most known ferrimagnetic compounds on Earth laboratories,such as HoFe \cite{8} we would have to shield some magnetic effects in the cylinder to be able to detect torsion effects on a torsion balance experiment. However since the presence of magnetic fields would be always present even for some superconductors our model seems to be a more realistic geometrical model that could be used in the LAB. Another possibility is to use this model to investigate models of spinning strings in superfluids and finding solutions of Einstein-Cartan theory for neutron stars with superfluid models. Part of this project have already begun with the investigation of phonon scattering in superfluids \cite{9} in torsioned spacetime where holonomy was computed.
\begin{flushleft}
{\large Acknowledgements}
\end{flushleft}
I would like to thank Professors H.Soleng and P.S.Letelier for helpful discussions on the subject of this paper.I also thank my daughter Maria Carolina Pires de Andrade for her patience while this work was carried out. Financial support from CNPq. and UERJ $(Prociencia)$ is gratefully acknowledged.


\newpage
\begin{thebibliography}{9}
\bibitem{1} L.C.Garcia de Andrade,Class. Quantum Gravity (2001) 18,3097.
\bibitem{2} M.L.Bedran and L.C.Garcia de Andrade,Prog.Theor.Phys.(1983)12,1583.
\bibitem{3} H.Soleng,Class. and Quant. Gravity 7,(1990),999.
\bibitem{4} C.L\"{a}mmerzahl,Phys.Lett. A 228 (1997)223.
\bibitem{5} H. Soleng, Gen. Rel. and Gravitation (1992) 24,1,111.
\bibitem{6} W. Arkuszewski, W. Kopczynski and V.N. Ponomariev, Comm. Math. Phys.(1975)45,183.
\bibitem{7} A. Beessange,Class. and Quantum Gravity (2000) 17,2509.
\bibitem{8} Li-Shing Hou and Wei-tou Ni,Rotatable torsion balance equivalence principle experiment for spin-polarised,Mod. Phys. Lett. A (2000). 
\bibitem{9} C. Furtado, Garcia de Andrade, A. M. Carvalho and F. Moraes, Phonon scattering in superfluids, in preparation.


\end{thebibliography}
\end{document}